\documentclass[aip,cha,showpacs,reprint,amsmath,amssymb,amsfonts,floatfix]{revtex4-1}
\usepackage{graphicx}
\usepackage{hyperref}

\begin{document}
\title{A consistent approach for the treatment of Fermi acceleration in time-dependent billiards}
\author{A. K. Karlis}
\email{akkarlis@gmail.com}
\affiliation{Department of Physics, University of Athens, GR-15771 Athens, Greece}
\affiliation{Physikalisches Institut, Universit\"at Heidelberg, Philosophenweg 12, 69120 Heidelberg,
Germany}
\author{F. K. Diakonos}
\email{fdiakono@phys.uoa.gr}
\affiliation{Department of Physics, University of Athens, GR-15771 Athens, Greece}
\author{V. Constantoudis}
\email{vconst@imel.demokritos.gr} 
\affiliation{Institute of Microelectronics, NCSR Demokritos, P.O. Box 60228, Attiki, Greece}
\date{\today}

\begin{abstract}
The standard description of Fermi acceleration, developing in a class of time-dependent billiards, is given in
terms of a diffusion process taking place in momentum space. Within this framework the evolution of the
probability density function (PDF) of the magnitude of particle velocities as a function of the number of
collisions $n$ is determined by the Fokker-Planck equation (FPE). In the literature the FPE is constructed by
identifying the transport coefficients with the ensemble averages of the change of the magnitude of particle
velocity and its square in the course of one collision. Although this treatment leads to the correct solution
after a sufficiently large number of collisions has been reached, the transient part of the evolution of the
PDF is not described. Moreover, in the case of the Fermi-Ulam model (FUM), if a stadanrd simplification is
employed, the solution of the FPE is even inconsistent with the values of the transport coefficients used for
its derivation. The goal of our work is to provide a self-consistent methodology for the treatment of Fermi
acceleration in time-dependent billiards. The proposed approach obviates any assumptions for
the continuity of the random process and the existence of the limits formally defining the transport
coefficients of the FPE. Specifically, we suggest, instead of the calculation of ensemble averages, the
derivation of the one-step transition probability function and the use of the Chapman-Kolmogorov forward
equation.  This approach is generic and can be applied to any time-dependent billiard for the treatment of
Fermi-acceleration. As a first step, we apply this methodology to the FUM, being the archetype of
time-dependent billiards to exhibit Fermi acceleration.
\end{abstract}
\maketitle

\begin{quotation}
Fermi acceleration, that is the increase of the mean energy of an ensemble of particles due to random
collisions off moving scatterers, is clearly one of the most interesting physical mechanisms linked to
time-dependent
billiards. Despite this fact, the standard approach in the literature for its analytical treatment can at best
describe Fermi acceleration in the asymptotic time limit. Herein, we propose a methodology,
which describes the evolution of Fermi acceleration at all times and, even more, obviates any unclear or
\textit{ad hoc} assumptions, which can lead to inconstencies or unreliable results. We exemplify
the proposed approach in the prototype of billiards exhibiting Fermi acceleration; The
Fermi-Ulam model. 
\end{quotation}

\section{Introduction}
More than 60 years ago, Fermi \cite{Fermi:1949} proposed an intuitive mechanism for the explanation of the
origin of the highly energetic cosmic ray particles and ever since it has been a subject of intense study. The
mechanism consists in the increase of the mean energy of particles as a result of random collisions with
moving scatterers. Soon after his seminal paper, his co-worker Ulam introduced a simple mechanical model for
testing Fermi's idea \cite{Ulam:1961}, known as the Fermi-Ulam model (FUM), linking for the first time Fermi
acceleration with the study of time-dependent billiards.

Since the introduction of the FUM, the standard description of Fermi acceleration developing in a class of
time-dependent billiards is given in terms of a diffusion process taking place in momentum space
\cite{Lieberman:1972,Lichtenberg:1980,Lichtenberg:1992}. Within this framework the evolution of the
probability density function (PDF) of the
magnitude of particle velocities as a function of the number of collisions $n$ is determined by the
Fokker-Planck equation (FPE). In the literature the FPE is constructed by identifying the transport
coefficients with the ensemble averages of the change of the magnitude of particle velocity and its square in
the course of one collision \cite{Lieberman:1972,Lichtenberg:1980,Lichtenberg:1992,Loskutov}. Although this
treatment leads to the correct
solution after a sufficiently large number of collisions has been reached, the transient part of the evolution
of the PDF is not described. Moreover, in the case of the FUM, if a standard simplification is employed
---known as the static wall approximation (SWA) or the simplified Fermi Ulam Model (SFUM)--- the
solution of the FPE is even inconsistent with the values of the transport coefficients used for its
derivation.

The aim of the work presented, is to provide a self-consistent methodology for the derivation of the PDF of
particle velocities for all times. The proposed approach obviates any assumptions for the continuity of the
random process and the existence of the limits formally defining the transport coefficients of the FPE.
Specifically, we suggest, instead of the calculation of ensemble averages, the derivation of the one-step
transition probability function (TPF) and the use of the Chapman-Kolmogorov (forward) equation (CKE).  This
approach
is generic and can be applied to any time-dependent billiard for the treatment of Fermi-acceleration. As a
first step, we apply this methodology to the FUM, being the archetype of time-dependent billiards to exhibit
Fermi acceleration. In this context, we show that the FPE reported in the literature
\cite{Lichtenberg:1992} describing the evolution of the PDF of the magnitude of particle velocities is not
valid, and that the observed agreement for
$n\gg1$ between the analytical and numerical results, in this case, should be regarded as accidental, i.e.
due to the validity of the central limit theorem (CLT).

\section{Statistical description of Fermi Acceleration}\label{JumpProcess}
Fermi acceleration developing in a time-dependent billiard can be described in terms of a stochastic process
taking place in the velocity space. Let $W(v, z)$ denote the probability of a particle being at the velocity
$z$ to perform a jump to velocity $v$ in the course of a single collision and $\rho(v,n\vert v',n')$ the
probability of a particle to possess velocity $v$ after $n$ collisions given that at the $n'$th collision it
had velocity $v'$. This jump process can be described by the following equation:
\begin{equation}
\label{master}
\rho\left(v,n\vert v',n' \right)= \int dz \rho\left(z,n-1\vert v',n'\right)
W\left(v,z\right)
\end{equation}
Eq.~(\ref{master}) is exact, on the condition that the process is Markovian. From a physical point of view,
this means that the probability of a particle to experience a velocity jump equal to $\Delta v$ upon the $n$th
collision depends only on the velocity it had at the previous step, i.e at the $n$th-1 collision.

\subsection{The Fokker-Planck approximation}
The standard approach in the literature for the determination of the asymptotic behaviour of the PDF of
particle velocities, is the approximation of the jump process with a diffusion process, described by the FPE
\cite{Lieberman:1972,Lichtenberg:1980,Lichtenberg:1992}. This approximation has also been used for the
analytical treatment of Fermi acceleration developing in higher-dimensional billiards, like the simplified
periodic Lorentz gas \cite{Loskutov}, i.e. the oscillating hard circular scatterers oscillate only in the
velocity space. An equation of the form of the FPE can be derived from Eq.~(\ref{master}) as follows
\cite{Risken}:

If we introduce $\Delta v\equiv v-z$, then the integrand in Eq.~(\ref{master}) can be rewritten as,
\begin{equation}
\begin{split}
 \label{master_b}
\rho\left(v,n\vert v',n' \right)=& \int d(\Delta v) \rho\left(v-\Delta v,n-1\vert v',n'\right)\\
&\times W\left(v-\Delta v+\Delta v,v-\Delta v\right).
\end{split}
\end{equation}
Expanding the distribution function $\rho\left(v,\Delta v,v',n'\right)$ and the transition probability
function (TPF)
$W(v;\Delta v)$ in a Taylor series yields,
\begin{equation}
\begin{split}
 \label{TaylorExpansion}
\rho\left(v,n\vert v',n' \right)=&\int d(\Delta v)  \sum\limits_{m=0}^\infty\frac{(-1)^m}{m!}(\Delta
v)^m\\ &\times\frac{\partial^m}{\partial v^m} \rho\left(v,n-1\vert v',n'\right)
W\left(v+\Delta v,v\right)
\end{split}
\end{equation}
Integrating now Eq.~(\ref{TaylorExpansion}) over $\Delta v$ we obtain,
\begin{equation}
\label{TaylorExpansionMoments}
\rho\left(v,n\vert v',n' \right)=\sum\limits_{m=0}^\infty\frac{(-1)^m}{m!}\frac{\partial^m}{\partial
v^m} M_m(v) \rho\left(v,n-1\vert v',n'\right),
\end{equation}
where $M_m(v)$ stands for the $m$th moment of the TPF, i.e. $$M_m(v)=\int (\Delta
v)^mW(v+\Delta v,z)d(\Delta v).$$
Therefore,
\begin{widetext}
\begin{equation}
 \label{TaylorExpansionFinalForm}
\rho\left(v,n\vert v',n' \right)-\rho\left(v,n-1\vert v',n'
\right)=\sum\limits_{m=1}^\infty 
\frac{(-1)^m}{m!}\frac{\partial^m}{\partial v^m}M_m(v)
\rho\left(v,n\vert
v',n'\right)
\end{equation}
\end{widetext}
By truncating the above series to the second order, and further by approximating the discrete derivative
$$\Delta_{k} \rho(v,n|v',n')=\left[\rho(v,n+k|v',n')-\rho(v,n|v',n')\right]/k,$$ $(k=1)$ with the continuous
derivative
$\partial\rho(v,n|v',n')/\partial n$, for $n\gg1$ one obtains an equation resembling the FPE.
\begin{equation}
\begin{split}
\label{FPE}
\frac{\partial}{\partial n}\rho\left(
v,n\vert v',n'\right)=&-\frac{\partial}{\partial v}\left[B\rho\left(
v,n\vert v',n'\right)\right]\\
&+\frac{1}{2} \frac{\partial^2}{\partial
v^2}\left[ D\rho\left(v,n\vert v',n'\right)\right],
\end{split}
\end{equation}
where the coefficient $B$ and $D$ is the ensemble average of the change of particle velocities and its
square, respectively, in one mapping period.

The approximations applied above for the construction of the FPE are valid on the condition
that only very small jumps are probable and further that the solution $\rho\left(v,n\vert v',n'\right)$
varies slowly with $v$ so that one can perform the expansion in a Taylor series. More
formally \cite{Kampen:2007}, we demand that there exists a $\delta>0$,
\begin{subequations}
\begin{align}
\label{FPE conditions}
 &W(z+\Delta z,z)\approx0,&~\text{for}~\vert \Delta z\vert>\delta\\
&\rho(v+\Delta v,n\vert v',n')\approx \rho(v,n\vert v',n'),&~\text{for}~\vert \Delta v\vert<\delta.
\end{align}
\end{subequations}

In the literature \cite{Lichtenberg:1992} the derivation of an FPE from Eq.~(\ref{master}) for the statistical
description of Fermi acceleration is carried out
on an \textit{ad hoc} basis. As a consequence, as shown in the following, it has produced contradictory
results. Moreover, by construction, the description of Fermi acceleration with a continuous stochastic
process, can at best describe the statistics only for $n\gg1$. Hence, a full description of FA in a
time-dependent billiard can only be given in the context of a jump process and consequently by
Eq.~(\ref{master}).

\subsection{A complete description: The Chapman-Kolmogorov equation}
The study of the transient statistics can only be accomplished by means of the Chapman-Kolmogorov equation,
i.e. Eq.~(\ref{master}). Assuming that initially particle velocities are distributed according to
$\rho(v,0)=\delta(v-z)$, Eq.~(\ref{master}) can be rewritten in respect with the one-step TPF.
$W(v,v')$ as,
\begin{equation}
 \label{CKE}
\rho(v_n,n|z,0)=\int\cdots\int W(v_n,v_{n-1})\cdots W(v_1,z) \mathbf{dv},
\end{equation}
where $\mathbf{dv}=\prod\limits_{i=1}^{n-1}dv_i$. The derivation of the one-step TPF can be achieved by
determining the PDF $p(\mathbf{q})$ of the variables $\mathbf{q}\equiv
\{x_i\}$ appearing in the dynamical equation defining the velocity of a particle after a collision with the
moving boundary of the time-dependent billiard, $v_n=f(v_{n-1},\mathbf{q})$. Then, the TPF is
\begin{equation}
 \label{TPD General}
W(v_n,v_{n-1})=\int p(\mathbf{q})\delta\left[v_n-f(\mathbf{q},v_{n-1})\right]\mathbf{dq}.
\end{equation}

If the resulting TPF is a function of the difference of velocities at successive steps
$W(v_n,v_{n-1})=W(v_n-v_{n-1})$, Eq.~(\ref{CKE}) can be easily solved in the Fourier space. Specifically, if
this condition is met, then by taking the Fourier transform of Eq.~(\ref{CKE}) we find,
\begin{equation}
 \label{FourierSpace}
\mathcal{F}\left[\tilde{\rho}(v,n\vert
z,0)\right]=\left(2\pi\right)^\frac{n-1}{2}e^{-ikz}\left\{\mathcal{F}\left[W(v)\right] \right\} ^n
, 
\end{equation}
where $\mathcal{F}=1/(\sqrt{2\pi})\int\limits_{-\infty}^{\infty}\exp[-ikv]dv$.

Moreover, in this case an approximate solution can be obtained directly in the velocity space, using the
saddlepoint approximation technique \cite{Goutis:1995}. Specifically, from Eq.~(\ref{CKE}), one can derive the
moment generating function
\begin{equation}
\begin{split}
 \label{moment generating function}
\phi(t,n)=&\int_{-\infty}^{\infty}e^{tx}\rho(v,n\vert
z,0)dv\\=&\left(\int_{-\infty}^{\infty}e^{tv}W(v)dv\right)^ne^{tz}
\end{split}
\end{equation}
 of the velocity PDF. To find the saddlepoint $\hat{t}(v,n)$, we solve the equation $\kappa'(t,n)=v$, where
$\kappa(t.n)=\log(\phi(t,n))$. Then, the PDF is approximately,
\begin{equation}
\begin{split}
 \label{saddlepoint approximation}
\rho(v,n\vert
z,0)\approx&\sqrt{\frac{1}{2\pi\kappa''(\hat{t}(v,n))}}\\&\times\exp\left[\kappa(\hat{t}(v,n))-\hat{t}(v,
n)v\right ] .
\end{split}
\end{equation}

In the following sections we implement the proposed methodology in the prototype of time-dependent billiards
exhibiting Fermi acceleration; The Fermi-Ulam model (FUM).

\section{Fermi acceleration in the stochastic simplified FUM}
\label{sfum}
The Fermi-Ulam model, originally proposed for testing the feasibility of gaining energy
through scattering off moving targets, i.e. Fermi acceleration, consists of one harmonically
oscillating and one fixed infinitely heavy hard wall and an ensemble of non-interacting particles bouncing
between them. Ever since, many different versions of the original model have been suggested and investigated,
such as variants of the FUM with dissipation
\cite{Luna:1990,Leonel:2007a,Leonel:2006,Leonel:2007b,Oliveira:2008}, different deterministic or random
drivings of the moving wall \cite{Karlis:2007,Leonel:2009} the quantum-mechanical version
\cite{Jose:1986,Visscher:1987,Makowski:1991,Razavy:1991,Jain:1993,Glasser:2009} and the, so called, bouncer
model
\cite{Pustylnikov:1978}, where a particle performs elastic \cite{Ladeira:2007} or inelastic
\cite{Holmes:1982,Everson:1986,Celaschi:1987,Mehta:1990,Leonel:2008,Kowalik,Livorati:2008} collisions with an
oscillating infinitely heavy platform under the influence of a gravitational field. Recently, a hybrid version
of the FUM and the bouncer model has also been investigated \cite{Leonel:2005,Ladeira:2007b}.

The equations defining the dynamics of the FUM are of implicit form with respect to the collision time, which
complicates numerical simulations and hinders an analytical treatment. A simplification
\cite{Lichtenberg:1992} ---known as the
\textsl{static wall approximation} (SWA) \cite{Karlis:2006,Karlis:2007}--  consists in treating the
oscillating wall as immobile,
located at its equilibrium position, yet allowing the transfer of momentum upon impact with a particle as if
the wall were harmonically oscillating. This simplification has become over the time
the standard approximation for studying the FUM \cite{Leonel:2004a}. The SWA speeds-up numerical simulations
and facilitates the
analytical treatment of the problem, while it has been generalized to higher-dimensional billiards with
time-dependent boundaries, such as the time-dependent Lorentz Gas \cite{Karlis:2007,Loskutov}.

Let us consider, without loss of generality, a FUM consisting of a fixed wall on the right and a moving wall
on the left, oscillating with frequency $\omega$. If we further assume that the positive direction of particle
velocities is towards the right, then the dynamics of the billiard within the framework of the SWA is defined
by the following set of dimensionless difference
equations:
\begin{subequations}
\label{dynamical_system}
\begin{eqnarray}
\label{eq1a}
t_n&=&t_{n-1} + \frac{2}{v_{n-1}} \\
\label{eq1b} v_n&=&|v_{n-1} + 2 u_n|\\
u_n&=&\epsilon \cos(t_n+\eta_n),
\end{eqnarray}
\end{subequations}
where $u_n$ is the velocity of the ``oscillating'' wall, $v_n$ is the algebraic value of the particle velocity
immediately \textit{after} the $n$th collision with the ``oscillating'' wall measured in units of $\omega w$
($w$ denoting the spacing between the walls), $t_n$ the time when the $n$th collision occurs measured in units
of $1/\omega$, $\eta_n$ a random variable uniformly distributed in the interval $[0,2\pi)$ updated immediately
after each collision between a particle and the fixed wall and $\epsilon$ the dimensionless ratio of the
amplitude of oscillation to the spacing between the ``oscillating'' and the fixed wall. It is noted that in
all numerical simulations $\epsilon$ was fixed at $1/10$.

The absolute value in  Eq.~({\ref{eq1b}}) is introduced in order to avoid the occurrence of positive particle
velocities after a collision with the ``oscillating'' wall, which would lead to the escape of the particle
from the area between the walls. It should be stressed that such a collision, within the framework of the
exact model, corresponds to a particle experiencing at least one second consecutive collision with the
``oscillating'' wall. Therefore, if $|V_{n-1}|<2|u_n|$ and $u_n\le0$, in order to prevent the particle from
escaping the region between the walls the velocity is reversed artificially. The presence of the absolute
value function in Eq.~(\ref{eq1b}), nevertheless, complicates the analytical treatment of the acceleration
problem. For this reason, it has become a standard practice in the treatment of the FUM to remove the absolute
value function, thereby neglecting the set of collision events upon which the particle direction is not
reversed after its collision with the ``oscillating'' wall. However, this further simplification gives rise to
a fundamental inconsistency: the ensemble mean of the absolute velocity obtained analytically does not change
through collisions with the ``{\it oscillating}'' wall, despite the well-established numerical result that
Fermi acceleration does take place in the phase-randomized FUM.

\subsection{The asymptotics of the PDF of particle velocities}
\subsubsection{Application of the central limit theorem}\label{CLT}
In this section we will discuss the asymptotic behaviour of the PDF of particle velocities in the SFUM.
Evidently, after $n$ collisions the velocity of a particle evolving in the SFUM is the sum of the velocity
jumps it has experienced up to this point, i.e. $v_n=\sum\limits_{m=1}^n\Delta v_n+v_0$.
Furthermore, due to Fermi-acceleration developing in the SFUM, after $n\gg1$ collisions, the vast majority of
the particles has acquired velocities much greater than the maximum wall velocity, irrespective of the
initial distribution. Therefore, most of the collisions, after a sufficiently large ``time'', take
place in the high velocity regime. In this limit, the absolute value function can be neglected and we
immediately obtain \cite{Lieberman:1972,Karlis:2008} $\langle\Delta v\rangle=0$ and $\left\langle(\Delta
v)^2\right\rangle=2\epsilon^2$. Therefore, the velocity jumps are completely uncorrelated, i.e. do not depend
on the velocity the particle had at the previous step. Thus, for $n\gg1$ and $v\gg\epsilon$ the central limit
theorem (CLT) dictates that the PDF of particle velocities tends to a Gaussian distribution, with a mean value
equal to $\sum\limits_{i=1}^n\langle\Delta v_i\rangle$ and variance
$\sigma^2=\sum\limits_{i=1}^n\left[\left\langle\left(\Delta v_i\right)^2\right\rangle-\left\langle\Delta
v_i\right\rangle^2\right]$.

\begin{figure}[htb!]
\includegraphics[width=8.6cm]{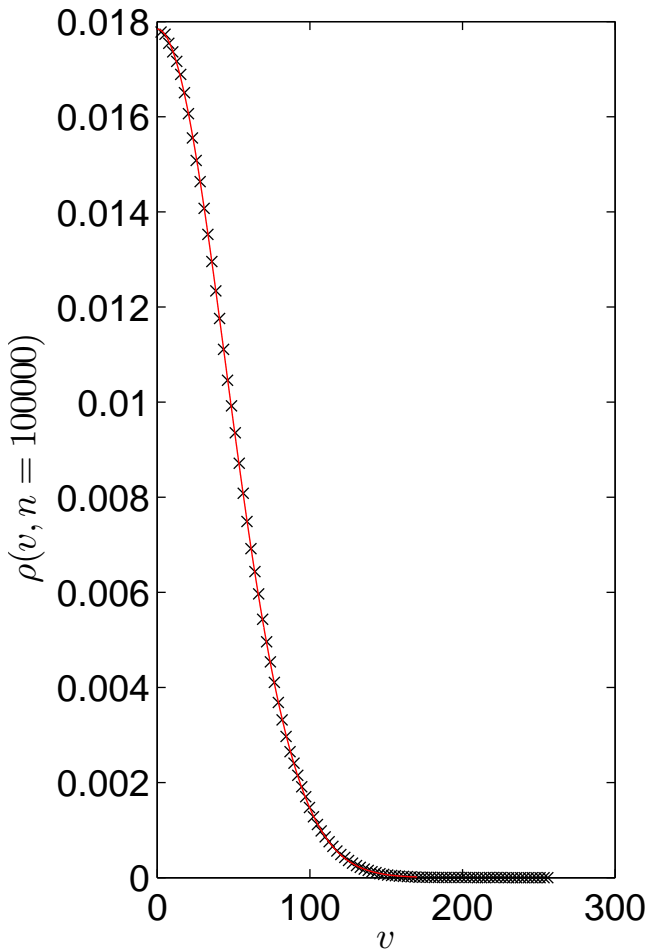}
\caption{Histogram ---diagonal crosses--- of particle velocities after $n=10^5$ collisions, obtained by the
iteration of Eqs.~(\ref{dynamical_system}), on the basis of an ensemble of $1.2\times10^6$ particles initially
distributed as $\rho(v,0)=\delta(v-\epsilon)$. The analytical result derived through the application of the
CLT [Eq.~(\ref{gaussian})] is also plotted ---solid (red) line. \label{CLTFig}}
\end{figure}

Hence, the PDF of particle velocities for $n\gg1$ is
\begin{equation}
 \label{gaussian}
\rho(v,n)=\frac{1}{\epsilon\sqrt{\pi n}}\exp\left[{-\frac{v^2}{4\epsilon^2 n}}\right].
\end{equation}
In Fig.~\ref{CLTFig}, Eq.~(\ref{gaussian}) is plotted along with the histogram of particle velocities obtained
from the simulation of $1.2\times10^6$ trajectories for $n=10^5$ collisions. The ensemble was initially
distributed according to the delta function $\delta(v-\epsilon)$. The analytical result obtained from the
application of the CLT is in perfect agreement with the numerically computed PDF.

\subsubsection{FPE equation in the SFUM}\label{FPE SFUM}
As mentioned in the previous section, assuming that for $n\gg1$ the probability measure of the events
occurring in the low-velocity regime is negligible, $B\equiv\langle \Delta v\rangle\backsimeq 0$,
$D\equiv\langle \left(\Delta v\right)^2\rangle\backsimeq 2\epsilon^2$.
In this limit Eq.~(\ref{FPE}) obtains the form of a standard diffusion equation. which for a delta initial
distribution of velocities $v=z$ together with reflecting conditions at
$v=0$ has as a solution the sum of two spreading Gaussians
\begin{equation}
\begin{split}
\label{FPESol}
\rho\left(v,n|z,0\right)=\frac{1}{2\sqrt{\pi n\epsilon^2}}\bigg\{&\exp\left[{-\frac{(v-z)^2}{4
n \epsilon^2}}\right]\\
&+\exp\left[{-\frac{(v+z)^2}{4n \epsilon^2}}\right]\bigg\},
\end{split}
\end{equation}
which for $n\ge z^2/(4\epsilon^2\ln2)$ transforms to Eq.~(\ref{gaussian}).

\subsubsection{Remarks}\label{CLT remarks}
Although the solution derived by means of the FPE is in agreement with the one obtained from the application
of the CLT, the methodology
used for the derivation of Eq.~(\ref{FPE}) stands on very shaky ground, since the termination of the series at
the second term in Eq.~(\ref{TaylorExpansionFinalForm}) is completely arbitrary\cite{Kampen:2007}. In general,
a jump process
can be approximated by a diffusion process, on the condition that a scaling assumption for the transition
probability holds. Namely, in the limit of infinitely small time intervals, the jumps should become smaller
and more frequent, such that the random process can be viewed as a continuous one \cite{Gardiner}. An
intuitive way to examine this is to consider the average square of the
jump size $\langle\left(\Delta
v\right)^2\rangle$ a particle makes having a velocity $v$ prior to the collision.

Given that the SWA treats the moving wall as fixed in the configuration space, all phases upon collision are
possible, independently of the velocity $v$ of particles before a collision. As a result, the average jump
size is not reduced as $v\to0$. In contrast, within the exact model, as the velocity of the particle prior
to a collision decreases, it becomes increasingly probable to collide with the wall at the turning points,
where its velocity is close to zero. Moreover, if the velocity of the particle before
a collision is small, then successive collisions are likely to occur, the exact dynamics result to higher
exit velocities. Consequently, successive collisions render small particle velocities improbable, as opposed
to the SFUM, where as shown in Sec.~\ref{CLT}, $v=0$ is the most probable velocity.

Summarizing, the application of the CLT for the determination of the long-time statistics is much more
straightforward and renders the solution of a differential equation redundant. More importantly, the
assumption of continuity of the stochastic process describing Fermi acceleration, which is essential for the
construction of an FPE, is not required.

\subsection{Short-time statistics in the SFUM}\label{CK}

From Eqs.~(\ref{dynamical_system}) the particle velocity after the $n$th collision given that it had velocity
$z$ is,
\begin{equation}
 \label{velRule}
 v=-z-2u-2\left(z+2u\right)\Theta\left(2u+z\right),
\end{equation}
where $\Theta(x)$ is the Heaviside unit-step function.

According to Eq.~(\ref{eq1b}) the wall velocity $u_n$ is determined by the phase $\xi_n\equiv t_n+\eta_n$ of
oscillation at the instant of the $n$th collision. Due to the fact that in the stochastic SFUM
the phase is randomly shifted through the addition of a random number $\eta_n$ ---distributed uniformly in
the interval $(0,2\pi)$--- after each collision, the oscillation phase $\xi_n$ is completely uncorrelated
between collisions, following a uniform distribution. Furthermore, given that in the context of the SFUM the
wall remains fixed in the configuration space, the wall velocity upon collision does not depend on the
velocity of the particle, therefore, $u_n$ and $v_{n-1}$ are also uncorrelated. From the fundamental
transformation law of probabilities the PDF of the wall velocity upon collision is,
\begin{equation}
 \label{wallPDF}
p(u)=\frac{1}{\pi\sqrt{\epsilon^2-u^2}}.
\end{equation}
For the single-step TPF $W(v,z)$ we can write,
\begin{equation}
 \label{tProb}
W(v,z)=\int_{-\epsilon}^{\epsilon} p(u)\delta\left[v-v(u,z)\right]du.
\end{equation}
Substituting Eqs.~(\ref{wallPDF}) and (\ref{velRule}) into
Eq.~(\ref{tProb}) we obtain after integrating over $u$,
\begin{equation}
\begin{split}
\label{transitionProbability}
W(v,z)=\frac{1}{\pi}\bigg[&\frac{\Theta(2\epsilon-v-z,2\epsilon-z)}{\sqrt{4\epsilon^2-(v+z)^2}}\\
&+\frac{\Theta(2\epsilon-v+z,2\epsilon+v-z)}{\sqrt{4\epsilon^2-(v-z)^2}}\bigg]
\end{split}
\end{equation}

\begin{figure}[htb!]
\includegraphics[width=8.6cm]{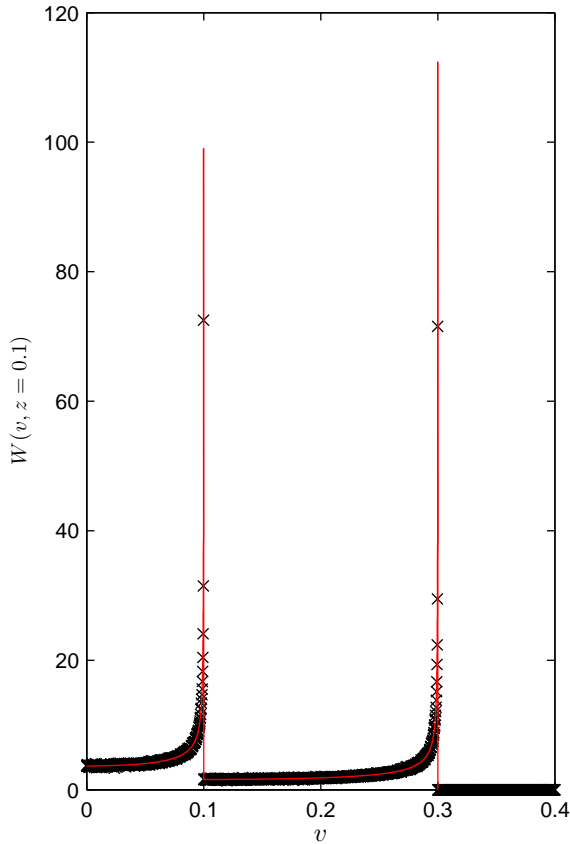}
\caption{Histogram ---diagonal crosses--- of particle velocities after a single collision with the ``moving''
wall, obtained using Eqs.~(\ref{dynamical_system}) and an ensemble of $1.2\times10^6$ particles with initial
velocity $z=0.1$. The analytical result [Eq.~(\ref{transitionProbability})] for the one-step transition
probability is also plotted for the sake of comparison ---solid line \label{transitionProbabilityFig}}
\end{figure}

In Fig.~\ref{transitionProbabilityFig} the analytical result of Eq.~(\ref{transitionProbability}) is compared
with the histogram of particle velocities after a single collision, obtained numerically using
Eqs.~(\ref{dynamical_system}) and an ensemble of $1.2\times10^6$ particles, with initial velocity
$z=0.1$. Clearly, the numerical and analytical results are in agreement.

The analytical result of Eq.~(\ref{transitionProbability}) reveals that the TPF depends
only on the most immediate history of a particle, that is on the velocity it had at the previous step.
Consequently, the stochastic process is indeed Markovian. Even more, if the particle before a collision has
velocity $z>2\epsilon$, then the velocity jump $\Delta v=v-z$ it undergoes is completely independent on its
history. Therefore, changes in velocity in the high-velocity regime are completely uncorrelated. 

In more detail, Eq.~(\ref{transitionProbability}), consists of two parts, one of which does indeed depend only
on the jump size.
However, the other branch of the TPF, taking effect for $v<2\epsilon$ ---relating to the
set of rare events\cite{Karlis:2008}--- depends also on the velocity at the last step. Nevertheless, the
action of both branches
of $W$ allows of a simple geometrical interpretation: At each step, the second branch of the TPF stretches the
PDF $\rho(v,n|z,0)$, resulting to a probability flux towards negative
values
of velocity. This unphysical result caused by the stretching is negated by the first branch, which folds the
part of the $\rho$
density over the vertical line at $v=0$. Therefore, the solution of Eq.~(\ref{CKE}) can be
obtained by
extending the domain of $\rho(v,n\vert z,0)$ to the whole real line and applying the method of images. Thus,
for any
number of collisions, we have 
\begin{equation}
 \label{methodOfImages}
\rho(v,n|z,0)=\tilde{\rho}(v,n|z,0)+\tilde{\rho}(v,n|-z,0),
\end{equation}
where $\tilde{\rho}$ is the solution of the unrestricted problem. Substituting
Eq.~(\ref{transitionProbability}) into Eq.~(\ref{FourierSpace}) we obtain,
\begin{equation}
 \label{FourierSpaceFinal}
\tilde{\rho}(k,n|z,0)=\frac{1}{\sqrt{2\pi}}\exp{(-ikz)}J_0\left(2\epsilon|k|\right)^n
\end{equation}
Eq.~(\ref{FourierSpaceFinal}) cannot be inverted analytically. To obtain an analytical result into the
velocity space, we use the saddlepoint approximation [Eq.~({\ref{saddlepoint approximation})]. The moment
generating function of $\rho(v,n\vert z,0)$ is,
\begin{equation}
 \label{MGF SFUM}
\phi(t,n)=\left(I_0(2 t \epsilon )\right)^ne^{tz},
\end{equation}
where $I_0$ is the modified Bessel function of the first kind. Consequently, the characteristic function is
$\kappa(t,n)=\log\phi(t,n)=n \log (I_0(2 t \epsilon ))+t z$. The
saddlepoint is the point $\hat{t}(v,n)$ that satisfies 
\begin{equation}
 \label{saddlepoint SFUM}
\kappa'(t,n)=v\Rightarrow \frac{2 n \epsilon  I_1(2 t \epsilon )}{I_0(2 t \epsilon )}+z=v.
\end{equation}
Eq.~(\ref{saddlepoint SFUM}) is implicit and cannot be solved analytically. To derive an explicit equation
we expand $\kappa'(t,n)$ in powers of $\epsilon$ to second order. Doing so we get,
\begin{equation}
 \label{that}
\hat{t}(v,n)=\frac{v-z}{2n\epsilon^2}.
\end{equation}
Substituting Eq.~(\ref{that}) into Eq.~(\ref{saddlepoint approximation}) we have,
\begin{widetext}
\begin{equation}
 \label{result of saddlepoint approximation}
\rho(v,n\vert z,0)\approx \frac{1}{2 \epsilon }e^{-\frac{(v-z)^2}{2 n \epsilon ^2}} I_0\left(\frac{v-z}{n
\epsilon }\right)^n\left[\frac{I_0\left(\frac{v-z}{n \epsilon }\right)^2}{\pi  n
I_0\left(\frac{v-z}{n
   \epsilon }\right)^2+\pi  n I_2\left(\frac{v-z}{n \epsilon }\right) I_0\left(\frac{v-z}{n
   \epsilon }\right)-2 \pi  n I_1\left(\frac{v-z}{n \epsilon }\right)^2}\right]^{1/2}.
\end{equation}
\end{widetext}

\begin{figure}[htb!]
\includegraphics[width=8.6cm]{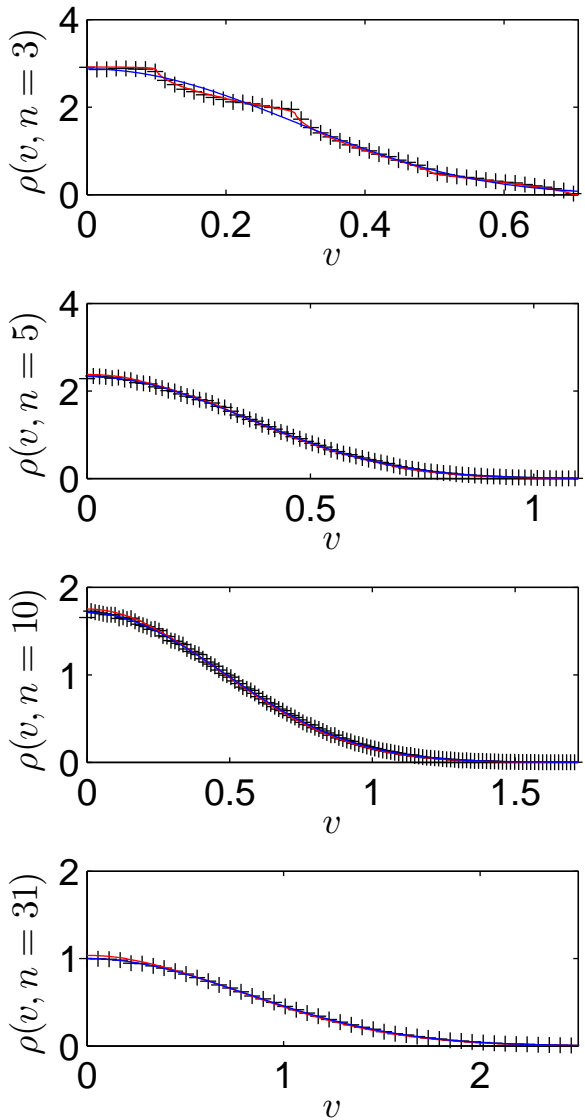}
\caption{Histogram ---upright crosses--- of particle velocities after $n=\{3,5,10,31\}$ collisions, obtained
by the
iteration of Eqs.~(\ref{dynamical_system}), on the basis of an ensemble of $1.2\times10^6$ particles initially
distributed as $\rho(v,0)=\delta(v-\epsilon)$. The exact numerical solution of Eq.~(\ref{CKE}) [red solid
line] as well as the
approximate one given by Eq.~(\ref{result of saddlepoint approximation}) [blue solid line] for
$n=\{3,5,10,31\}$, using only the second branch of the one-step TPF [Eq.~(\ref{transitionProbability})],
followed by the application of the method of images [Eq.~(\ref{methodOfImages})] are also plotted for the
sake of comparison. line.\label{CKS}}
\end{figure}

In Fig.~\ref{CKS} we present the exact numerical solution of Eq.~(\ref{CKE}) [red solid line] as well as the
approximate one given by Eq.~(\ref{result of saddlepoint approximation}) [blue solid line] for
$n=\{3,5,10,31\}$, using only the second branch of the one-step TPF [Eq.~(\ref{transitionProbability})],
followed by the application of the method of images [Eq.~(\ref{methodOfImages})]. The numerical solution is in
total agreement with the histogram of particle velocities obtained by the iteration of the dynamical equations
(\ref{dynamical_system}) [upright crosses], for all times. Even more, we see that the saddlepoint
approximative solution describes very accurately the evolution of the PDF for $n\ge5$. As can be observed, the
PDF of particle velocities quickly approaches to a Gaussian distribution, in accordance with the prediction of
the CLT. This is attributed to the fact that the TPF can be reduced to a difference kernel. Consequently, the
additional assumption we made for the application of the CLT in Sec.~\ref{CLT}, namely that the statistical
weight of the rare events\cite{Karlis:2008} is negligible, is redundant. This can be circumvented, as
aforementioned, by extending the domain of particle velocities. Thus, if one applies the CLT on the whole real
line, then all the conditions for its applications are met exactly. As a final remark, we would like to stress
that the success of the Fokker-Planck type of equation reported in the literature for even short times is
attributed to the validity of the CLT, guaranteeing that the PDF will converge to a normal distribution,
allowing for the use of a diffusion equation. If however, the reduction of the TPF to a difference kernel is
not feasible, then the transient can be arbitrarily long, a point demonstrated via an example in the following
section.

\subsection{Long Transients}\label{modifiedSFUM}
In the last section we showed that the specific choice made for treating negative velocities after a
collision, i.e. reflection with respect to the $v=0$, reduces the TPF to an even function of the jump size.
As a consequence, the PDF of particle
velocities approaches rapidly to a sum of two spreading Gaussians. Clearly, after a number of collisions the
system will ``forget'' its initial distribution, and the sum will converge to a single half-Gaussian centered
at $v=0^+$. Therefore, the most probable velocity for a particle evolving in the phase-randomized SFUM will
eventually be $v_p=0^+$, in clear contrast with the results given by the numerical simulation and analytical
results derived using the exact dynamical mapping \cite{Karlis:2006,Karlis:2007}, according to which as
$v\rightarrow 0$, $\rho(v,n)\rightarrow 0$. From a physical point of view this happens because if the motion
of the wall in the configuration space is taken into account, as $v\rightarrow 0$ collisions resulting in an
energy loss can occur only in a small neighborhood around the wall's extreme positions, where its velocity is
zero, resulting to a minimal energy loss. Furthermore, if the particle velocity is comparable to the wall
velocity, consecutive collisions can take place, resulting in a higher exit velocity from the interaction
region within the exact model.

On account of these properties of the collision process in the exact model, the reflection of negative
velocities is not realistic. To gap the difference between the results of the simplified and the exact FUM, we
propose instead of the inversion of negative particle velocities, the inversion of the direction of the
wall's velocity, if the collision would lead to a negative particle velocity. This would lead in a greater
energy gain in comparison with the reflection, as $|v+u|\le|v|+|u|$. Therefore, Eqs~(\ref{dynamical_system})
change to,
\begin{subequations}
\label{dynamical_system_2}
\begin{eqnarray}
\label{eq1aa}
t_n&=&t_{n-1} + \frac{2}{V_{n-1}} \\
\label{eq1bb} V_n&=&V_{n-1} + 2 |u_n|\\
u_n&=&\epsilon \cos(t_n+\eta_n).
\end{eqnarray}
\end{subequations}

Let us now derive the TPF. From Eqs.~(\ref{wallPDF}), (\ref{tProb}) and
(\ref{eq1bb}) we obtain,
\begin{widetext}
\begin{equation}
\label{transitionProbabilityB}
W(v,z)=\frac{\Theta(2\epsilon+z-v)}{\pi\sqrt{4\epsilon^2-(v-z)^2}}\big\{
\Theta(2\epsilon+z-v)+\Theta(\epsilon-z/2)\left[\Theta(z-v-2\epsilon)+\Theta(v-2z)\right]\big\}
\end{equation}
\end{widetext}

A comparison of the analytical result given by Eq.~(\ref{transitionProbabilityB}) and the histogram
of velocities obtained on the basis of an ensemble of $1.2\times10^6$ particles after 1 iteration of
Eqs.~(\ref{dynamical_system_2}) is presented in Fig~.\ref{FigTPb}, proving the validity of the
derived result.

\begin{figure}[htb!]
\includegraphics[width=8.6cm]{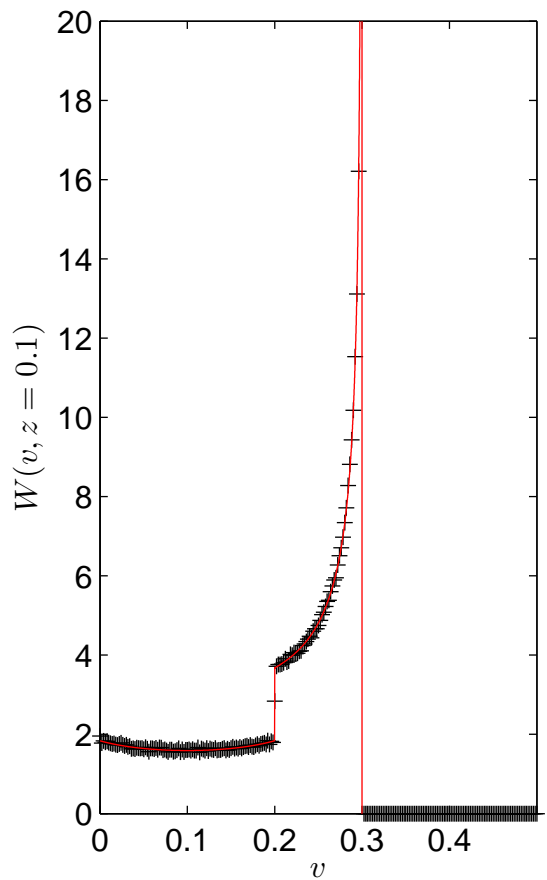}
\caption{Histogram ---diagonal crosses--- of particle velocities after a single collision with the ``moving''
wall, obtained using Eqs.~(\ref{dynamical_system_2}) and an ensemble of $1.2\times10^6$ particles with initial
velocity $z=0.1$. The analytical result [Eq.~(\ref{transitionProbabilityB})] for the one-step transition
probability is also plotted for the sake of comparison ---solid line \label{FigTPb}}
\end{figure}

As expected, the TPF has two branches, one taking effect only in the low velocity regime,
i.e. $z<2\epsilon$ and another which is nonzero for any velocity $z$ prior to a collision. As was also the
case in the SFUM with reflection of negative velocities, the part of the TPF that is
relevant to the low-velocity regime, depends on the jump size, as well as, on the velocity of
a particle prior to a collision. However, due to the fact that this branch of $W$ does not have a
simple geometrical interpretation, the single-step transition function cannot be reduced to a difference
propagator by an extension of the domain of $W$ to the whole real line. Thus, the conditions for the
application of the CLT are not met exactly. However, due to the acceleration of the particles, as
$n\rightarrow \infty$, the probability measure of particles having velocity $z<2\epsilon$ becomes negligible.
Therefore, for $n\gg1$ and $v\gg\epsilon$ the PDF of particle velocities tends to a Gaussian
distribution [Eq.~(\ref{gaussian})].

\begin{figure}[htb!]
\includegraphics[width=8.6cm]{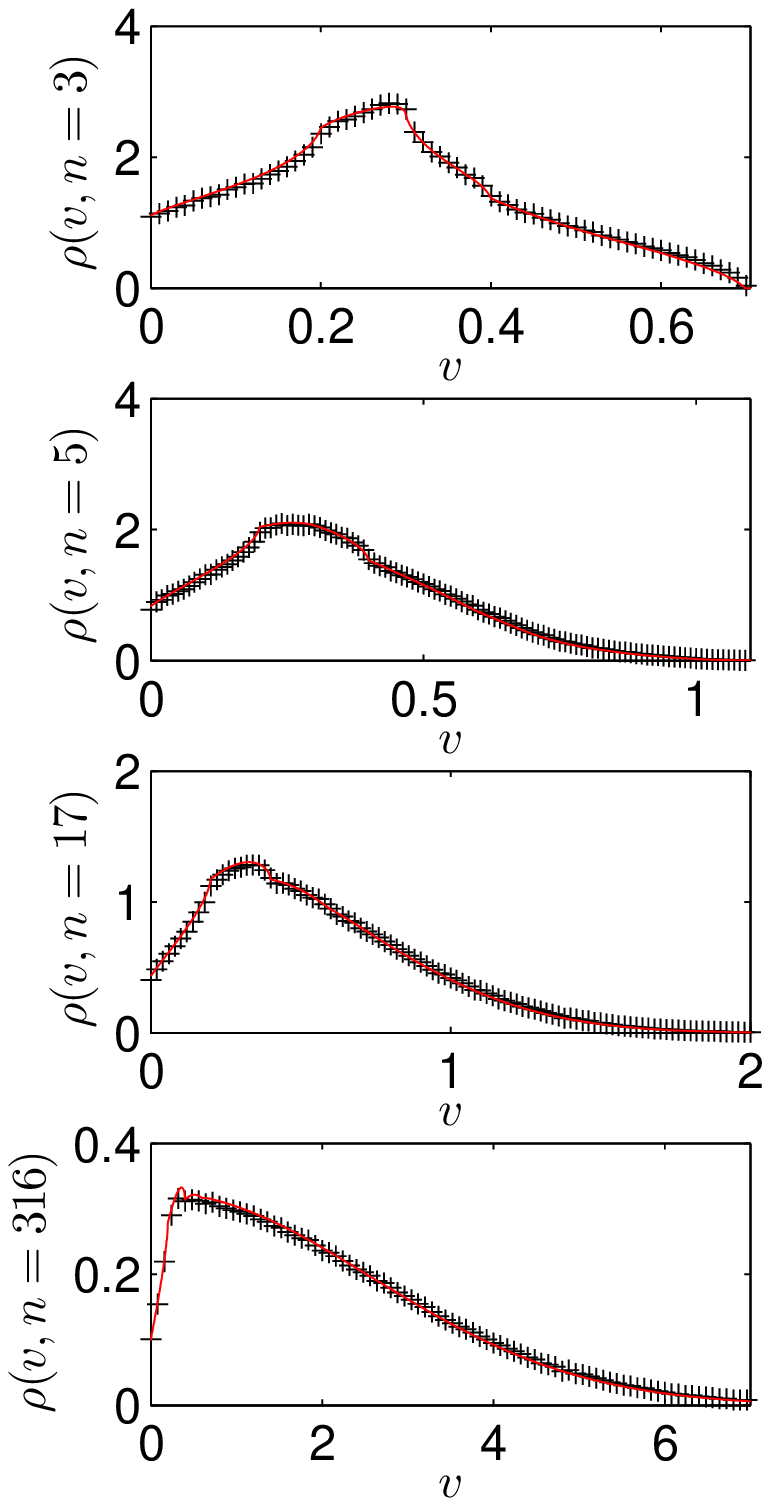}
\caption{Histogram ---upright crosses--- of particle velocities after $n=\{3,5,17,316\}$ collisions, obtained
by the
iteration of Eqs.~(\ref{dynamical_system_2}), on the basis of an ensemble of $1.2\times10^6$ particles
initially
distributed as $\rho(v,0)=\delta(v-\epsilon)$. The solution obtained by numerically solving the forward CKE
[Eq.\ref{CKE}] using as the TPF of the modified version of the SFUM
[Eq.~\ref{transitionProbabilityB}] is also plotted ---solid (red) line. \label{CKSB}}
\end{figure}

The study of the transient behaviour of the PDF requires the solution of the CKE [Eq.~(\ref{CKE})]. The
numerical solution of Eq.~(\ref{CKE}) at times $n=\{3,5,17,316\}$ is presented in Fig.~\ref{CKSB}. The
histograms of particle velocities for the same times, calculated by iterating an ensemble of $1.2\times10^6$
particles for up to $n=10^5$ collisions, are also plotted for the sake of comparison. It can be seen, that the
solution of the CKE is in agreement with the results of the simulation for all times presented. In
Fig.~\ref{figCLT2} the histogram of velocities for $n=10^5$ collisions is plotted. The solution obtained from
the application of the CLT ---on the assumption that the statistical weight of collisions happening in the
region $v<2\epsilon$ is negligible, is also plotted, and is full agreement with the PDF in this velocity
region. However, a blow-up of the low-velocity region shows that even after $10^5$ collisions, the PDF
diverges from the Gaussian profile. This is clear evidence that even after a very large number of collisions,
the PDF in the whole velocity domain cannot be described by an FPE, in contrast to the standard version of the
SFUM [Eqs.~(\ref{dynamical_system})] even though the argumentation used in both cases was the same, i.e. the
particles are accelerated. This exemplifies the potential pitfalls of a diffusion approximation of
Fermi-acceleration in time-dependent billiards.

\begin{figure}[htb!]
\includegraphics[width=8.6cm]{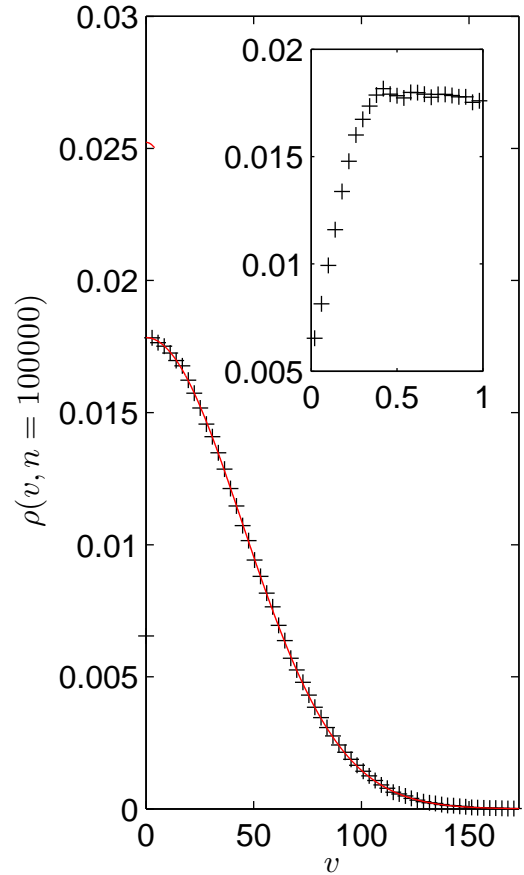}
\caption{Histogram ---upright crosses--- of particle velocities after $n=10^5$ collisions, obtained
by the
iteration of Eqs.~(\ref{dynamical_system_2}), on the basis of an ensemble of $1.2\times10^6$ particles
initially
distributed as $\rho(v,0)=\delta(v-\epsilon)$. The asymptotic Gaussian distribution [Eq.~(\ref{gaussian})]
predicted by the CLT is also plotted --solid (red) line. A blow-up of the numerically obtained
histogram at the low-velocity region is illustrated in the inset.\label{figCLT2}}
\end{figure}

\subsection{Fermi acceleration in the exact FUM}\label{exact}
The simplification employed to the treatment of Fermi acceleration in the FUM ---treating the wall as fixed
in real space--- however widespread and thoroughly studied, prohibits the study of
the details of Fermi acceleration. In Refs.~\cite{Karlis:2006,Karlis:2007,Karlis:2008} it was proved
that small additional fluctuations of the time of collision due to dynamical correlations induced by the
displacement of the scatterer upon impact, quantitatively as well as qualitatively change the evolution of
the PDF of velocities, increasing the efficiency of Fermi acceleration. Moreover, the development of
correlations causes the CLT to break down and the asymptotic PDF ceases to be a normal distribution.

In Ref.~\cite{Karlis:2006} utilizing a novel simplification, the so-called hopping approximation, which
succeeds into retaining all the essentials of the exact dynamics, an analytical solution for the asymptotic
behaviour of the PDF of particle velocities in the exact FUM, which was in excellent agreement with the
numerical simulation of the exact FUM. Specifically, it was shown by means of a Fokker-Planck type of equation
that, in contrast with the SFUM, the attractor of the PDF of velocities in the function space is a
Maxwell-Boltzman like distribution, i.e. independently of the initial distribution of velocities, the PDF
converges to a Maxwell-Boltzman like distribution. Therefore, in the case of the exact FUM for $v\to 0^+$,
$\rho(v,n\vert z,0)\to 0$, in contrast to the SFUM where $\rho(v,n\vert z,0)$ attains its maximum
value for $v\to0^+$. This difference between the simplified and the exact FUM can be understood as follows: If
the velocity of a particle after a collision with the moving wall is small, then multiple successive
collisions are likely to occur within the exact FUM, resulting into higher exit velocities, as opposed to the
simplified model, within which successive collisions cannot be realized.
 
Another subtle difference between the simplified and the exact FUM [see Sec.~\ref{CLT
remarks}] is that within the exact FUM particles with low velocity are more likely to collide with the
oscillating wall near its turning points, where the wall velocity is close to 0. Hence, the velocity jump
performed by a particle due to a collision with the wall $\Delta v\to 0$ as $v\to 0$. As a result, Fermi
acceleration when using the exact dynamics can be better approximated by a continuous stochastic
process, or equivalently, by the FPE. Still, the transient statistics in the system can only be studied by
means of the CKE.

As aforementioned, the movement of the wall in the configuration space described by the exact dynamics
 results into a more efficient energy transfer from the moving wall to the
particles upon collision, when compared to the SFUM. In mathematical terms, this causes the PDF of the
oscillation phase on collisions
to deviate from the uniform distribution, reflecting the fact that head-on collisions are more preferable
than head-tail collisions. However, the phase of oscillation of the moving wall when a particles collides
with the fixed wall ---or when it passes through any fixed point within the area between the two walls
comprising the FUM--- is uniformly distributed. The map describing the exact dynamics is,
\begin{subequations}
\label{exact_dynamics}
\begin{align}
d_{n,}&=\epsilon\sin\left(\delta
t_n+t_{n-1}+\eta_n\right)\\
u_n&=\epsilon\cos(\delta t_n + t_{n-1} + \eta_n)\\
v_n&=v_{n-1} + 2 u_n,
\end{align}
\end{subequations}
where $d_{n,}$ stands for the position of the moving wall
in the instant of the $n$th collision, $u_n$ for the wall
velocity, $\eta_n$ for the random phase component and $v_n$ for the
particle velocity after the $n$th collision. The time of free flight
$\delta t_n$ is obtained by solving the implicit equation
\begin{equation}
\label{eq:eq2}x_{n-1} + v_{n-1} \delta t_n= d_{n},
\end{equation}
where $x_n$ stands for the position of the particle in the instant
of the $n$th collision.

\begin{figure}[htb!]
\includegraphics[width=8.6cm]{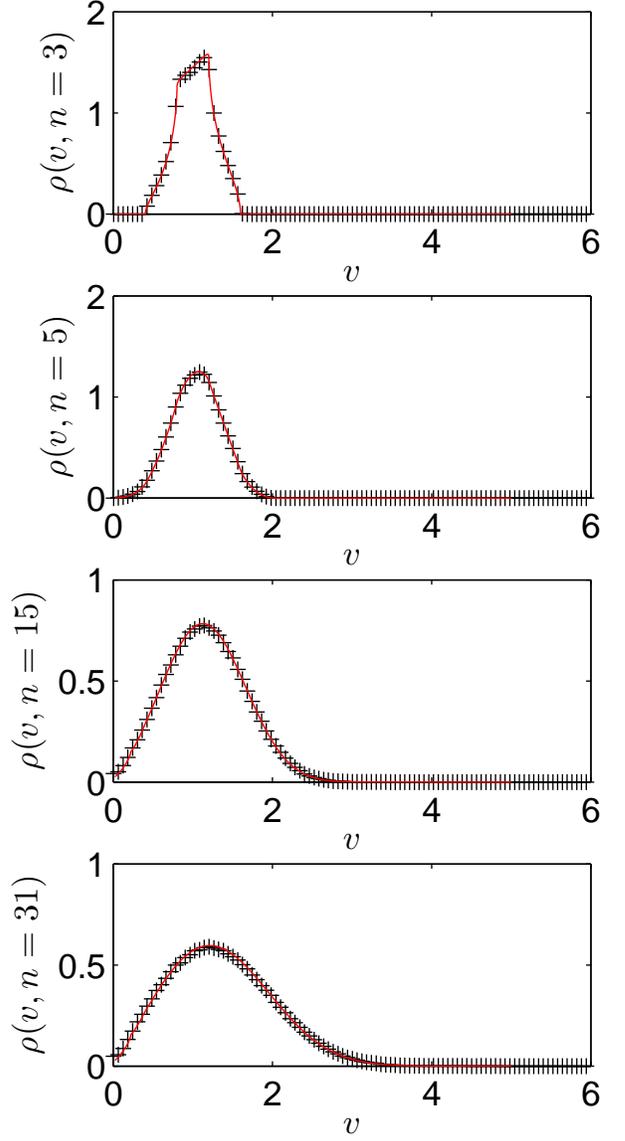}
\caption{Histogram ---upright crosses--- of particle velocities after $n=\{3,5,17,316\}$ collisions, obtained
by the
iteration of Eqs.~(\ref{exact_dynamics}), on the basis of an ensemble of $1.2\times10^6$ particles
initially
distributed as $\rho(v,0)=\delta(v-10)$. The solution obtained by numerically solving the forward CKE
[Eq.\ref{CKE}] using as the TPF of the exact model
[Eq.~\ref{transitionProbabilityExact}] is also plotted ---solid (red) line. \label{CKSExact}}
\end{figure}

If we denote the phase of oscillation of the moving wall when a particle collides with the fixed wall with
$\psi$, then
\begin{equation}
 \label{psiandxi}
\psi_n=\cos^{-1}\left(\frac{u_n}{\epsilon}\right)+\frac{1}{z}\left(1+\sqrt{\epsilon^2-u_n^2}\right)
\end{equation}
For Eq.~(\ref{psiandxi}) we obtain for the distribution of the wall velocity upon collision,
\begin{equation}
 \label{PDFWallVelocityExact}
p_e(u)=\frac{u+z}{\pi z\sqrt{\epsilon^2-u^2}}.
\end{equation}
Installing Eq.~(\ref{PDFWallVelocityExact}) into Eq.~(\ref{tProb}) we obtain the one-step TPF 
for the exact model
\begin{equation}
 \label{transitionProbabilityExact}
W_e(v,z)=\frac{\Theta(2\epsilon-|v-z|)(v+z)}{2\pi z\sqrt{4\epsilon^2-(v-z)^2}}.
\end{equation}
Inserting Eq.~(\ref{transitionProbabilityExact}) into the CKE we can numerically compute the evolution of the
PDF of particle velocities. In Fig.~\ref{CKSExact}, the numerical solution of the CKE is compared with the
histogram of particle velocities, obtained by simulating $1.2\times10^6$ trajectories using
Eqs.~(\ref{exact_dynamics}). The particles were initially distributed according to $\rho(v,0)=\delta(v-10)$.
Once more, the solution of the CKE is in complete agreement with the results of the simulation, proving that
the method can also be successfully employed when the exact dynamics are taken into account.

\section{Summary and conclusions}
Fermi acceleration, is one of the most interesting aspects of time-dependent billiards, as it has been
understood over the years, that it is a fundamental acceleration mechanism, playing a key role in a variety
of phenomena, far beyond its original scope, i.e. cosmic-ray particle acceleration.

Until now, the investigation of Fermi acceleration ---in the class of time-dependent billiards in which it
develops--- has been carried out via its approximation with a diffusion process. Within, this framework, the
evolution of the density of particle velocities was determined by a Fokker-Planck equation. However,
its derivation is always based on assumptions and approximations that rarely can be justified. Moreover,
its prediction power is limited in the long-term statistics of the system and no information is given for the
transient behaviour. Even more, its use in the SFUM, which is the first system that was successfully
investigated with the use of the FPE, is completely redundant, as the CLT yields the same
results in a far more straightforward manner.

Herein, we proposed a consistent methodology, which obviates unclear assumptions and even more, can give an
accurate description of the transient evolution of particle velocities. The cornerstone of this methodology is
the use of the Chapman-Kolmogorov equation. The fundamental difference in comparison with the traditional
approach using the FPE, is that no assumption for the continuity of the stochastic process describing Fermi
acceleration needs not to be made. Another advantage of the proposed approach is that all collision events can
be taken into account, which cannot be done in the construction of the FPE, and even when possible, it can
lead to less accurate results.

The method was successfully applied to the FUM, which is the prototype of time-dependent
billiards exhibiting Fermi acceleration. In specific, we studied the standard SFUM,
within which collisions leading to a potential escape from the system are handled by artificially inverting
the particle exit velocity, as well as a variant of the SFUM, where if a collision would lead to a particle
still moving towards the wall, the velocity of the wall is inverted before the collision takes place. Finally,
we showed how this method can be applied to the exact model, showing how the effect of the motion of the wall
in the configuration space can be included in the description of Fermi acceleration through the CKE. In all
three cases, the CKE yielded accurate results for all times. As a final remark, we would like to stress that
this methodology can be applied to higher-dimensional billiards \cite{Karlis:2011} and therefore is generic.

\section*{Acknowledgments}
This research has been co-financed by the European Union (European Social Fund –
ESF) and Greek national funds through the Operational Program "Education and Lifelong Learning" of the
National Strategic Reference Framework (NSRF) - Research Funding Program: Heracleitus II. Investing in
knowledge society through the European Social Fund.

This work was made possible by the facilities of the Shared Hierarchical 
Academic Research Computing Network (SHARCNET:www.sharcnet.ca) and Compute/Calcul Canada.


\begin{thebibliography}{99}
\bibitem{Fermi:1949} E. Fermi, Phys. Rev. \textbf{75}, 1169 (1949).
\bibitem{Ulam:1961} S. Ulam, in \textit{Proceedings of the Fourth Berkeley Symposium on Mathematics,
Statistics,
and Probability} (California University Press, Berkeley, 1961), Vol. 3, p. 315.
\bibitem{Lieberman:1972} M.A. Lieberman, A.J. Lichtenberg, Phys. Rev. A \textbf{5}, 1852 (1972).
\bibitem{Lichtenberg:1980} A.J. Lichtenberg, M.A. Lieberman, R.H, Cohen, Physica D \textbf{1}, 291 (1980).
\bibitem{Lichtenberg:1992} A.J. Lichtenberg, M.A. Lieberman, \textit{Regular and Chaotic Dynamics} (Springer
Verlag, New York, 1992) Vol. 38.
\bibitem{Loskutov} A.Yu. Loskutov, A.B. Ryabov, L.G. Akinshin, J. Exp. Theor. Phys. \textbf{89}, 966 (1999);
J. Phys. A: Math. Gen. 33, 7973 (2000).
\bibitem{Risken} H. Risken, \textit{The Fokker-Planck Equation, Methods of Solution and Applications}
(Springer-Verlag, Berlin, Heidelberg 1989) Vol. 18.
\bibitem{Kampen:2007} N. G. van Kampen, \textit{Stochastic Processes in Physics and Chemistry}, (Elsevier,
Amsterdam, 2007).
\bibitem{Goutis:1995} C. Goutis and G. Casella, \textit{Statistics and Econometrics Series 27}, Working Paper
95-63 (1995).
\bibitem{Luna:1990} G.A. Luna-Acosta, Phys. Rev. A \textbf{42}, 7155 (1990).
\bibitem{Leonel:2007a} E.D. Leonel, R.E. de Carvalho, Phys. Lett. A \textbf{364}, 475 (2007).
\bibitem{Leonel:2006}
E.D. Leonel, P.V.E. McClintock, Phys. Rev. E \textbf{73} 066223 (2006).
\bibitem{Leonel:2007b}  E.D. Leonel, J. Phys. A: Math. Theor. \textbf{40}, F1077 (2007).
\bibitem{Oliveira:2008} D.F.M. Oliveira, E.D. Leonel, Braz. J. Phys. \textbf{38}, 62 (2008).
\bibitem{Karlis:2007} A.K. Karlis, P. K. Papachristou, F.K. Diakonos, V. Constantoudis, P. Schmelcher, Phys.
Rev. E \textbf{76}, 016214 (2007).
\bibitem{Leonel:2009} E.D. Leonel, E.P. Marinho, Physica A \textbf{388}, 4927 (2009).
\bibitem{Jose:1986}  J.V. Jos\'{e}, R. Cordery 1986 Phys. Rev. Lett. \textbf{56}, 290 (1986).
\bibitem{Visscher:1987}  W.M. Visscher, Phys. Rev. A \textbf{36}, 5031 (1987).
\bibitem{Makowski:1991} A.J. Makowski, S.T. Dembi\'{n}ski , Phys. Lett. A \textbf{154}, 217 (1991).
\bibitem{Razavy:1991} M. Razavy, Phys. Rev. A \textbf{44}, 2384 (1991).
\bibitem{Jain:1993} S. R. Jain, Phys. Rev. Lett. \textbf{70}, 3553 (1993).
\bibitem{Glasser:2009} M. L. Glasser, J. Mateo, J. Negro and L. M. Nieto, Chaos, Solitons and Fractals
\textbf{41}, 2067 (2009)
\bibitem{Pustylnikov:1978} L.D. Pustylnikov, Trans. Moscow Math. Soc. \textbf{2}, 1 (1978).
\bibitem{Ladeira:2007} D.G. Ladeira, J.K. da Silva, J. Phys. A: Math. Theor. \textbf{40}, 11467 (2007).
\bibitem{Holmes:1982} P.J. Holmes, Journal of Sound and
Vibration 84, 173 (1982).
\bibitem{Everson:1986} R.M. Everson, Physica D \textbf{19}, 355 (1986).
\bibitem{Celaschi:1987} S. Celaschi, R.L. Zimmerman, Phys. Lett. A \textbf{120}, 447 (1987).
\bibitem{Kowalik} Z.J. Kowalik, M. Franaszek, P. Piera\'{n}ski, Phys. Rev. A \textbf{37}, 4016 (1988).
\bibitem{Mehta:1990} A. Mehta, J.M. Luck, Phys. Rev. Lett. \textbf{65}, 393 (1990).
\bibitem{Leonel:2008} E.D. Leonel, A.L.P. Livorati, Physica A \textbf{387}, 1155 (2008).
\bibitem{Livorati:2008} A.L.P. Livorati, D.G. Ladeira, E.D. Leonel, Phys. Rev. E \textbf{78}, 056205 (2008).
\bibitem{Leonel:2005} E.D. Leonel, P.V.E. McClintock, J. Phys. A: Math. Gen. \textbf{38}, 823 (2005).
\bibitem{Ladeira:2007b} D.G. Ladeira, E.D. Leonel, Chaos \textbf{17}, 013119 (2007).
\bibitem{Karlis:2006} A.K. Karlis, P.K. Papachristou, F.K. Diakonos, V. Constantoudis,
P. Schmelcher, Phys. Rev. Lett. \textbf{97}, 194102 (2006).
\bibitem{Leonel:2004a} E.D. Leonel, P.V.E. McClintock and J.K. da Silva, Phys. Rev. Lett. \textbf{93}, 014101
(2004).
\bibitem{Karlis:2008} A.K. Karlis, F.K. Diakonos, V. Constantoudis, P. Schmelcher, Phys. Rev. E \textbf{78}, 
046213 (2008).
\bibitem{Gardiner} G.W. Gardiner, \textit{Handbook of stochastic Methods, for Physics, Chemistry
and the
Natural Sciences} (Springer Verlag, New York, 1985).
\bibitem{Karlis:2011} A. K. Karlis, F. K. Diakonos and V. Constantoudis, in preparation.
\end{thebibliography}
\end{document}